\documentclass[a4paper]{article}

\usepackage{icrc2013}
\usepackage[english]{babel}

\title{Modeling the time and energy behavior of the GCR intensity\\ in the periods
of low activity around the last three solar minima}

\shorttitle{Modeling GCR intensity around the last three solar minima}

\authors{
Krainev M.B., Bazilevskaya G.A., Kalinin M.S., Svirzhevskaya A.K., Svirzhevsky N.S.
}

\afiliations{
Lebedev Physical Institute, Moscow, Russia}

\email{mkrainev46@mail.ru}

\abstract{Using the simple model for the description of the GCR modulation in the heliosphere and the sets of parameters discussed in the accompanying paper we model some features of the time and energy behavior of the GCR intensity near the Earth observed during periods of low solar activity around three last solar minima. In order to understand the mechanisms underlying these features in the GCR behavior, we use the suggested earlier decomposition of the calculated intensity into the partial intensities corresponding to the main processes (diffusion, adiabatic losses, convection and drifts).
}

\keywords{modeling GCR intensity, GCR intensity around solar minima, unusual solar minimum 23/24}

\begin{document}
\maketitle

%Begin a section.
\section{Introduction}
The period of low solar activity of the last solar cycle (SC) 23 was rather strange not only because of the record--setting heliospheric and GCR characteristics in the minimum 23/24 between SC 23 and 24 (see references in our accompanying papers \cite{Bazilevskaya_etal_ICRC33_0274_2013,Kalinin_etal_ICRC33_0297_2013}). Some other time and energy details were also unusual \cite{Svirzhevsky_etal_ICRC31_icrc1105_2009,McDonald_etal_GRL_37_L18101_2010,Mewaldt_etal_ApJL_723_L1-L6_2010,bib:baz,Bazilevskaya_etal_CosmicResearch_51(1)_29-36_2013}: the abrupt change of the energy dependence of the GCR modulation some time before the moments of the maximum of the intensity; the unusual sequence of these moments for the low and high energy particles; the unusual correlation between changes of the different heliospheric parameters. A few works were devoted to the interpretation of these details
\cite{Kota_Jokipii_ICRC32_11_12-14_2011}.

In this paper using the simple model for the description of the GCR modulation and the sets of parameters discussed in \cite{Kalinin_etal_ICRC33_0297_2013} we try to reproduce some of the above time and energy features in the GCR intensity near the Earth around three last solar minima. Besides, in order to understand the mechanisms underlying these features in the GCR behavior, we discuss the behavior of some other GCR characteristics, the radial gradients and the partial intensities corresponding to the main processes (diffusion, adiabatic losses, convection and drifts), which we  calculate using the method of decomposition of the calculated intensity suggested in \cite{Krainev_Kalinin_JoPCS_409_012155_2013,Krainev_Kalinin_BulRAS_77(5)_510-512_2013}.

\section{The model}
In \cite{Kalinin_etal_ICRC33_0297_2013} we discuss the differential boundary--value problem for the distribution function $U(\vec r,p,t)=J(\vec r,T,t)/p^2$ \cite{Parker_PhysRev_110_1445_1958,Krymsky_GaA_4_977_1964,Jokipii_Levy_Hubbard_ApJ_213_861_1977}:
\begin{equation}
	\textrm{\hspace{-3mm}}\underbrace{- \nabla\cdot ({\cal K}\nabla U)}_{\mbox{diff.}}+\underbrace{{\vec{V}}^{sw}\nabla U}_{\mbox{convect.}}\underbrace{-\frac{\nabla\cdot{\vec{V}}^{sw}}3p\frac{\partial U}{\partial p}}_{\mbox{adiab.loss}}+\underbrace{{\vec{V}}^{dr}\nabla U}_{\mbox{drift}}=0 \label{TPE}
\end{equation}
with the usual boundary conditions at $r=r_{min},r_{max}$ and poles (without termination shock and heliosheath) and the "initial" condition $\left.U\right|_{p=p_{max}}=U_{um}(p_{max})$, where $J$, $p$ and $T$ are the intensity, momentum and kinetic energy of the particles, $p_{max}=150$ GeV/c and $U_{um}$ is the distribution function of the unmodulated GCRs. The set of constant model parameters $\left\{\eta_i\right\}_c$ is chosen in \cite{Kalinin_etal_ICRC33_0297_2013} as well as the sets of the main $\left\{B_{r,E},\alpha_t,V_{sw,E} \right\}_m$ and additional $\left\{r_{max},\alpha_R\right\}_a$ parameters necessary to describe the GCR intensity in the minima of the last solar cycles. Here $B_{r,E},V_{sw,E},\alpha_t$ are the magnitude of the HMF radial component, the solar wind velocity (both near the Earth) and the heliospheric current sheet (HCS) tilt angle, respectively, and $r_{min}$,  $r_{max}$ and $\alpha_R$ are the minimum and maximum radii of the modulation region and the index in the rigidity dependence of the GCR diffusion coefficients, respectively.

So using the same $\left\{\eta_i\right\}_c$, the measured values of $\left\{B_{r,E},V_{sw,E},\alpha_t \right\}_m$  and the additional factors $\left\{r_{max},\alpha_R\right\}_a$ one can calculate the GCR intensity for any time in the whole heliosphere for any energy. Here we are interested in comparison of the time and energy behavior of the calculated and observed GCR proton intensity near the Earth ($r=r_E=1$ AU, $\vartheta=\vartheta_E=90$ deg ) in three periods of low solar activity around solar activity minima 21/22, 22/23 and 23/24 for the low $T_{low}=200$ MeV and high $T_{high}=15000$ MeV energy particles.

We compare the observed GCR intensity with that calculated using the main heliospheric characteristics averaged for one year before the moments when the GCR intensity was observed, since this is the characteristic time for the solar wind to reach the heliospheric boundary \cite{Potgieter_etal_Apxiv_1302.1284_2013}. As the periods of low solar activity we consider the periods with the "dipole" type of the HMF polarity distribution (with the only and global HCS, see \cite{Krainev_Kalinin_etal_ICRC33_0317_2013}) when the quasi--tilt $\alpha_{qt}$, the half of the heliolatitude HCS range calculated using the Wilcox Solar Observatory (WSO) model \cite{WSO_Site}, can be used as a HCS tilt angle $\alpha_t$ in the models of drift velocity \cite{Jokipii_Thomas_ApJ_243_1115_1981,Huttings_Burger_ASR_16_9_1995,
Kalinin_Krainev_ECRS21_222-225_2009}.

\section{Additional calculated GCR characteristics}
In the process of solving boundary--value problem (\ref{TPE}) for each time run (the Carrington rotation) the finite difference approximation of the radial gradient of the relative intensity $u(r_E,\vartheta_E,T)\equiv J(r_E,\vartheta_E,T)/J_{um}(T)$ is saved for each step in energy and then its time behavior can be considered for the low and high energy particles.

The decomposition of the calculated intensity into the partial ``intensities'' connected with the main processes of the GCR modulation is discussed in more details in \cite{Krainev_Kalinin_JoPCS_409_012155_2013,Krainev_Kalinin_BulRAS_77(5)_510-512_2013}. Here we only mention that just as the radial gradient of the relative intensity in the process of solving boundary--value problem  for each step in energy we also save the finite difference approximation of each term of (\ref{TPE}) for $r_E,\vartheta_E$. Then we reconsider the usual partial transport equation (\ref{TPE}) as the ordinary differential equation with respect to momentum which can be easily integrated as the approximation of all terms have been already memorized for all energies. In such a way (and converting from the distribution function to intensity) we get the partial ``intensities'' corresponding to the diffusion $J_p^{diff}$, convection $J_p^{conv}$ and drift $J_p^{drift}$, and their sum $J^{dcd}$.
As we also know the total calculated intensity $J$, we can calculate the difference $J-J^{dcd}$ and call it the partial ``intensity'', corresponding to the adiabatic loss, $J_p^{adiab}$. So we get $J=J_p^{diff}+J_p^{conv}+J_p^{adiab}+J_p^{drift}$. As this decomposition is made for each time run we can study the time behavior of each partial "intensity" along with the total calculated intensity for the low and high energy particles.

The inverted commas in the ``intensities'' emphasize the conventionality of these terms, which can be negative if the process leads to the reduction of the total intensity. Of course, the meaning of, for example, the partial diffusion intensity is not simply the contribution of this process in the total intensity, as the diffusion term in (\ref{TPE}) depends on the intensity gradient which is the product of all processes.

\section{General features of the GCR intensity}
\begin{figure}[h]
  \centering
  \includegraphics[width=0.5\textwidth]{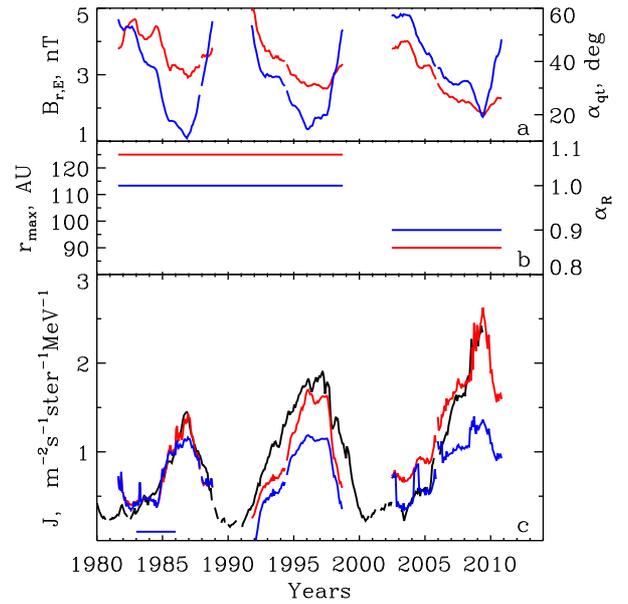}
  \caption{The time--dependent parameters of the model and the GCR intensity in 1980--2013.
  In the panels: (a) the absolute value of the HMF radial component (red) and the HCS quasi--tilt (blue);
  (b) the radius of the modulation region (red) and the index of the rigidity dependence of the parallel diffusion coefficient (blue); (c) the observed (black) and calculated (red) low energy GCR intensity and also the calculated high energy intensity (blue) normalized to the calculated low energy intensity by the linear regression for the period 1983--1986, shown by the horizontal blue line near the time axis. The observed $J_{low}(t)$ is shifted by half a year back in time.
  }
  \label{fig1}
 \end{figure}

In Fig.\ref{fig1} the time profiles of some time--dependent parameters of the model and of the GCR intensity near the Earth (both calculated and observed) are shown for the periods around three last solar minima. The 27d averaged HMF characteristics near the Earth \cite{OMNI_Site} and the HCS quasi--tilt $\alpha_{qt}$ (\cite{WSO_Site}, classic) are yearly smoothed. As the observed low energy proton GCR intensity $J(200$ MeV) we use for 1973--2006 $J_p(120-230$ MeV) (IMP8/GME, reported in \cite{McDonald_SpaceScience_ser_ISSI_33_1998}) and for 2006--2009  $J_p(110-240$ MeV) (PAMELA, constructed in \cite{Bazilevskaya_etal_ICRC33_0274_2013} from the data reported in \cite{Adriani_etal_ApJ_765_2_91_2013}). The time behavior of the calculated high energy proton GCR intensity $J(15000$ MeV) is the proxy of the neutron monitor data (Moscow, effective energy $T_{eff}\approx 15000$ GeV).

As to the behavior of the main time--dependent parameters of the model one can see the gradual (almost linear) change from solar minimum 21/22 through 22/23 to 23/24 opposite for $B_{r,E}$ and $\alpha_{qt}$ which was discussed in \cite{Kalinin_etal_ICRC33_0297_2013}. The time profile of $V_{sw,E}$ is not shown in Fig.\ref{fig1} but it demonstrates the same gradual decrease for the last three minima. For the calculations in this paper we decided not to change the additional parameters $\left\{r_{max},\alpha_R\right\}_a$ around each solar minimum, while their change between minima 21/22--22/23 and 23/24, also discussed in \cite{Kalinin_etal_ICRC33_0297_2013} is easily seen in Fig.\ref{fig1} (b).

The calculated low energy GCR intensity (red line in Fig.\ref{fig1}(c)) is very close to the observed one for almost the whole period around 21/22 solar minimum (except for 1981--1982) and also for 2008--2009 before the 23/24 solar minimum. Note that the HMF polarity $A$ is negative for both of these periods. In contrast, for the A-positive period around the 22/23 solar minimum the time profile of the calculated GCR intensity is more narrow than that of the observed intensity although the well--known tendency of the alternating peak-like (for $A<0$) and more flat (for $A>0$) proton intensity time profiles is seen in the calculated GCR intensity as well.

Since the calculated high energy GCR intensity is shown for all three periods as normalized to the calculated low energy intensity with the single normalization period 1983--1986 before the intensity maximum in 1987, it is only natural that both curves almost coincide around 21/22 minimum (with small but distinct softening of the calculated GCR spectrum in 1986--1987). However, the decrease of the calculated high energy normalized intensity with respect to the calculated low  energy one is very distinct for the whole $A$--positive period around 22/23 solar minimum and for the first half of $A$--negative period before solar minimum 23/24 getting even greater for 2008--2009. So the results of the calculation support the conclusion of \cite{Bazilevskaya_etal_ICRC33_0274_2013} on the gradual softening of the GCR variation spectrum when one goes from 21/22 to 22/23 solar minima getting exceptional near 23/24 minimum. Note, however, that softening of the GCR spectrum during the A--positive period (1990--2000) with respect to the A--negative one (1980--1990) is to some extend just the manifestation of the magnetic cycle, as the cross--over of the differential spectra for these two periods occurs at $T_{co}\approx 10$ GeV (see \cite{Kalinin_etal_ICRC33_0297_2013}).

\section{Additional features of the GCR intensity}
\begin{figure}[h]
  \centering
  \includegraphics[width=0.5\textwidth]{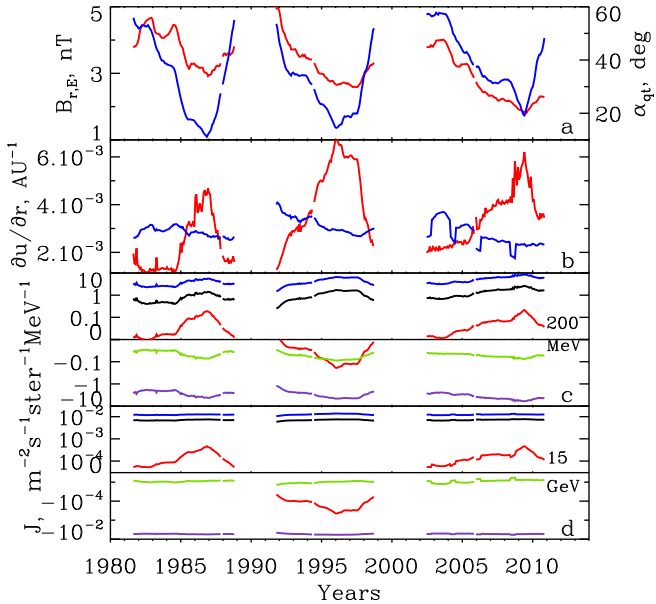}
  \caption{The detailed GCR characteristics in 1980--2013. In the panels:
  (a) the absolute value of the HMF radial component (red) and the HCS tilt (blue);
  (b) the calculated radial gradient of the relative intensity near the Earth for the low energy (red) and high energy (blue) particles;
  (c) the total calculated GCR intensity (black) and the partial ``intensities'' for the low energy: connected with the diffusion (blue), convection (green), adiabatic loss (violet) and drift (red);
  (d) the same as in panel (c) but for the high energy particles.
  Note that the intensities in panels (c) and (d) are shown in the LOG--scale but taking into account their sign.
  }
  \label{fig2}
 \end{figure}

In Fig.\ref{fig2} beside the total GCR intensity, the time profiles of the additional GCR characteristics (the radial gradients of the relative intensity and partial ``intensities'' connected with the main processes of the modulation) are shown for the low and high energy particles along with time profiles of $B_{r,E}$ and $\alpha_{qt}$.

The time behavior of the calculated local relative intensity gradients is rather interesting and unexpected. First, why does the radial gradient increase when one approaches the moment of the intensity's maximum for the low energy particle for both types of HMF polarity, while for the high energy particles the situation is opposite? Its behavior for the high energies looks as more expected as we are accustomed to the idea that the radial gradients decrease as one goes from maximum to minimum of the solar cycle. Second, for the ``normal'' pair of solar minima, 21/22 and 22/23, the calculated radial gradients of the intensity are higher for A-positive than for A-negative period, although we are accustomed to the opposite behavior of the radial gradient. However, in this case one should remember that this usual behavior concerns the radial gradients in the free heliosphere far from both its outer and inner regions while the gradients and intensities shown  in Fig.\ref{fig2} (b) are for the inner heliosphere, rather near the Sun with its strong magnetic fields. And third, why does the radial gradient of the high energy intensity manifest such a strong and abrupt quasi--periodical variations (something like quasi--biannual ones) while for the low energy particles (and even for the heliospheric modulating factors) these variations are much less evident? Now we cannot answer all these questions, but the answers are very important for understanding the GCR intensity behavior and especially for the partial ``intensities'' as the latter are directly connected to the intensity gradients by their definition.

 \begin{figure}[h]
  \centering
  \includegraphics[width=0.5\textwidth]{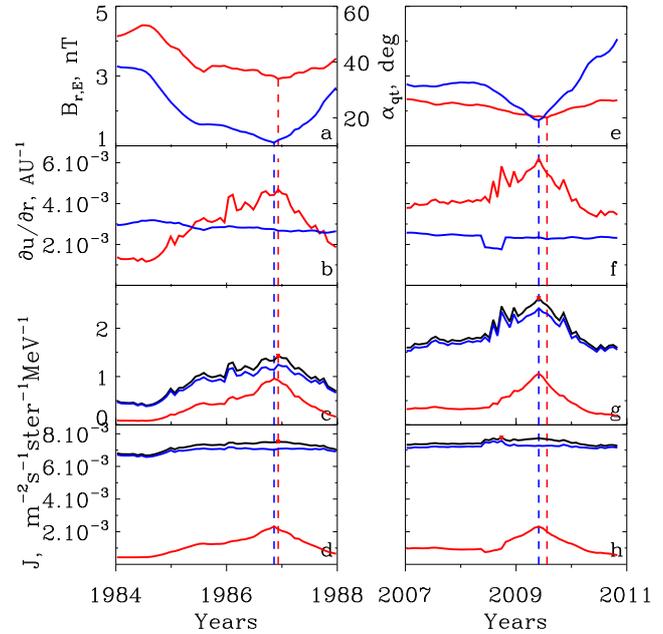}
  \caption{The detailed GCR characteristics in 1984--1988 (left) and 2007--2011 (right).
  In the panels: (a, e) the absolute value of the HMF radial component (red) and the HCS quasi--tilt (blue). The moments of their minima are shown by the vertical dashed lines of the corresponding color;
  (b, f) the calculated radial gradient of the relative intensity near the Earth for the low energy (red) and high energy (blue) particles;
  (c, g) the total calculated GCR intensity (black) and the main partial ``intensities'' for the low energy: the sum of the partial ``intensities'' connected with the diffusion, convection and adiabatic loss (blue) and drift (multiplied by 5, red);
  (d, h) the same as in panel (c) but for the high energy particles.
  }
  \label{fig3}
 \end{figure}

The behavior of the calculated partial ``intensities'' is also interesting. First, it can be seen that the diffusion partial intensity is always positive while the convective and adiabatic ones are always negative. It means that for the case considered (the inner heliosphere near the equator around solar minima) the diffusion always increases the intensity while the convection and adiabatic losses decrease it. The drift partial intensity is positive for A--negative and negative for A--positive periods. As to their magnitude for the case considered the diffusion term is the greatest while the convection term is the smallest (their ratio is about 100). The magnitude of the drift term is intermediate and its relative contribution increases as the tilt diminishes. It is interesting that near the solar minimum the magnitude of the drift term for A--negative period is of the same magnitude as for A--positive periods and, as can be easily shown, the drift term almost entirely consists of the current--sheet drift component. It means that the flat form of the GCR intensity time profile during A--positive solar minima can be due to the fact that the pointed in magnitude drift contribution {\underline{reduces}} the intensity making the profile V--like (or at least more flat), while for A--negative periods it {\underline{enhances}} the intensity making the profile $\Lambda$--like. Note that this conclusion is opposite to the usual view that the GCR intensity time profile is flat around the $qA>0$ solar minimum because in this case the particles come to the Earth from the high latitudes and don't feel the form of the HCS, \cite{Jokipii_Thomas_ApJ_243_1115_1981}

The adiabatic term is almost of the same magnitude as the diffusion term but of the opposite sign. So it can be useful to construct their sum, then add to it also small convection term calling the result the diffusion--adiabatic--convection (dac) partial GCR intensity $J_p^{dac}$. Then the total intensity can be considered as the sum of two main partial ``intensities'', $J=J_p^{dac}+J_p^{drift}$. In Fig.\ref{fig3} along with the main heliospheric modulating factors and the radial gradients of the relative intensity the time profiles of the above main GCR partial intensities are compared for 4--year periods around two last solar minima with the same HMF polarity $A<0$. Note that now to see the details better we used the linear scale for the intensities and the drift partial intensity is multiplied by 5.

The first interesting detail one can see from Fig.\ref{fig3} is that for the low energy particles the magnitudes of the drift term for 21/22 (1987) and 23/24 (2009) solar minima are about the same (the lower HMF strength in 2009 is compensated by the higher tilt angle) while the dac--term and total intensity are significantly (50--70 \%) higher for the second minimum. The situation with the high energies looks much more symmetric, that is the small excess of the total intensity in 2009 consists of comparable excesses in both dac-- and drift terms. It means that according to our model the additional flux of low energy particles in 2009 with respect to 1987 which is discussed in \cite{Svirzhevsky_etal_ICRC31_icrc1105_2009,bib:baz,Kalinin_etal_ICRC33_0297_2013} is due mainly to enhanced diffusion rather than drift. The second interesting detail is that the time profile of the calculated intensity is much more pointed in 2009 than in 1987 and the moment of its maximum in 2009 coincides with the moment of minimum of tilt, not of HMF strength, as in 1987. Finally the third feature seen in the relative behavior of different main partial intensities is the change of the drift contribution when compared with the diffusion (or dac--term), especially sudden for 23/24 period. In the middle of 2008 the drift term started to grow faster while the dac--term almost stopped growing. Probably this feature can be relevant to the change in 2008 of the  observed GCR intensity variation spectrum reported in \cite{bib:baz}.

Note that in this paper we have not used all the potentials of the model. We have not tried to imitate the change of $B/\delta B$ in correlation with the HCS tilt reported in \cite{Bazilevskaya_etal_CosmicResearch_51(1)_29-36_2013}, $B$ and $\delta B$ being the HMF strength and its variation, respectively. Besides, the gradual component of the change of  $r_{max}\propto \sqrt{N_{sw}V_{sw}^2}$ in 1980--2000 could result in the gradual softening of the GCR intensity variation in accordance with the observations \cite{Bazilevskaya_etal_ICRC33_0274_2013}. At least this explanation of the gradual softening of the GCR intensity variation would look more substantiated by the observations than the gradual change of the energy dependence of the diffusion coefficient which was used in \cite{Potgieter_etal_Apxiv_1302.1284_2013} to reproduce the PAMELA energy spectrum in 2006--2009 \cite{Adriani_etal_ApJ_765_2_91_2013}.

\section{Conclusions}
1. Using rather simple model of the GCR modulation in the heliosphere it is possible to reproduce to some extend the important features of the time and energy behavior of the GCR intensity in the periods around the last three solar activity minima: the general form of the time profiles of the low energy GCR intensity;
the great excess of the low energy GCR intensity during the last solar minimum; the gradual softening of the GCR spectrum when one goes from one period around solar minimum to the next one.

\noindent 2. To understand the mechanisms underlying the observed features of the time and energy behavior of the GCR intensity in the periods around the last three solar activity minima the use of some additional calculated characteristics of the GCR intensity (the local intensity radial gradients and partial ``intensities'' connected with the main processes of the modulation) can be very useful.

\noindent 3. The relative changes of the different partial ``intensities'', probably, indicate to the causes of some peculiarities in the GCR intensity observations (what mechanisms are behind the different forms of the intensity--time profiles during periods of opposite HMF polarity; what mechanism is mainly responsible for the energy dependent excess of the GCR intensity during the last solar minimum; how the change of the different heliospheric parameters influences the different components of the GCR intensity etc). Some features of the behavior of the radial gradients of the GCR intensity are still intriguing and need further consideration.

\vspace*{0.5cm}
\footnotesize{{\bf Acknowledgment:}{ We thank the Russian Foundation for Basic Research (grants 11-02-00095a, 12-02-00215a, 13-02-00585a, 13-02-10006k) and the Program ``Fundamental Properties of Matter and Astrophysics'' of the Presidium of the Russian Academy of Sciences.}}


\begin{thebibliography}{}

\bibitem{Bazilevskaya_etal_ICRC33_0274_2013}
G.A. Bazilevskaya et al., icrc2013-0274.pdf (2013)

\bibitem{Kalinin_etal_ICRC33_0297_2013}
M.S. Kalinin et al., icrc2013-0297.pdf (2013)

\bibitem{Svirzhevsky_etal_ICRC31_icrc1105_2009}
N.S. Svirzhevsky et al, Proc. 31st ICRC (2009) \verb| http://icrc2009.uni.lodz.pl/proc/pdf/icrc1105.pdf|

\bibitem{McDonald_etal_GRL_37_L18101_2010}
F.B. McDonald et al., Geophys. Res. Lett. 37 (18) (2010) L18101 doi:10.1029/2010GL044218.

\bibitem{Mewaldt_etal_ApJL_723_L1-L6_2010}
R.A. Mewaldt et al., Astrophys. J. Lett. 723 (2010) L1-L6 doi:10.1088/2041-8205/723/1/L1.

\bibitem{bib:baz} G. A. Bazilevskaya et al., Advances in Space Research 49(4) (2012)  784-790 doi:10.1016/j.asr.2011.12.002.

\bibitem{Bazilevskaya_etal_CosmicResearch_51(1)_29-36_2013}
G. A. Bazilevskaya et al., Cosmic Research 51 (2012) 29-36.

\bibitem{Kota_Jokipii_ICRC32_11_12-14_2011}
J. Kota, J.R. Jokipii, Proc. 32rd ICRC 11 (2011) 12-14.

\bibitem{Krainev_Kalinin_JoPCS_409_012155_2013}
M.B. Krainev, M.S. Kalinin, J. of Physics: Conf.Series 409 (2013) 012155.

\bibitem{Krainev_Kalinin_BulRAS_77(5)_510-512_2013}
M.B. Krainev, M.S. Kalinin, Bulletine of Russian Academy of Sciences Physics 77 (2013) 510-512.

\bibitem{Parker_PhysRev_110_1445_1958}
E.N. Parker, Phys. Rev. 110 (1958) 1445.

\bibitem{Krymsky_GaA_4_977_1964}
G.F. Krymsky, Geomagnetism and Aeronomy 4 (1964) 977-985 (in Russian)

\bibitem{Jokipii_Levy_Hubbard_ApJ_213_861_1977}
J.R. Jokipii et al., Ap. J. 213 (1977) 861--861.

\bibitem{Potgieter_etal_Apxiv_1302.1284_2013}
M.S. Potgieter et al., Solar Physics Apxiv-1302.1284.pdf (2013)

\bibitem{Krainev_Kalinin_etal_ICRC33_0317_2013}
M.B. Krainev and M.S. Kalinin, icrc2013-0317.pdf (2013)

\bibitem{WSO_Site}
\verb|http://wso.stanford.edu/|

\bibitem{Jokipii_Thomas_ApJ_243_1115_1981}
J.R. Jokipii and B. Thomas B, Ap.J. 243 (1981) 1115-1122.

\bibitem{Huttings_Burger_ASR_16_9_1995}
M. Hattingh, R.A. Burger, Adv. Space Sci. 16(9) (1995) 9.

\bibitem{Kalinin_Krainev_ECRS21_222-225_2009}
M.S. Kalinin, M.B. Krainev, Proc. 21st ECRS (2008) 222-225.

\bibitem{OMNI_Site}
\verb|ftp://omniweb.gsfc.nasa.gov/pub/data/omni|

\bibitem{McDonald_SpaceScience_ser_ISSI_33_1998}
F.B. McDonald, in Cosmic Rays in the Heliosphere (eds.) Fisk L A et al. Space Science series of ISSI (1998) 33-50.

\bibitem{Adriani_etal_ApJ_765_2_91_2013}
O. Adriani et al., Ap. J. 765(2) (2013) 91.

\end{thebibliography}
\end{document}